\documentstyle[mprocl]{article}

\def\Journal#1#2#3#4{{#1} {\bf #2}, #3 (#4)}

\def\NPB{{\em Nucl. Phys.} B}
\def\PLB{{\em Phys. Lett.}  B}
\def\PRL{\em Phys. Rev. Lett.}
\def\PRD{{\em Phys. Rev.} D}
\def\ZPC{{\em Z. Phys.} C}

\def\be{\begin{equation}}
\def\ee{\end{equation}}
\def\bea{\begin{eqnarray}}
\def\eea{\end{eqnarray}}

\begin{document}

\title{ON SOFT SUSY BREAKING PARAMETERS
IN STRING MODELS WITH ANOMALOUS $U(1)$ SYMMETRY}

\author{Yoshiharu KAWAMURA}

\address{Department of Physics, Shinshu University \\
Matsumoto, 390-0802 Japan 
\\E-mail: haru@azusa.shinshu-u.ac.jp} 


\maketitle\abstracts{We study the magnitudes of soft SUSY
breaking parameters
in heterotic string models with anomalous $U(1)$ symmetry. 
In most cases, $D$-term contribution to soft scalar masses 
is expected to be comparable to or dominant over other contributions
provided that supersymmetry breaking
is mediated by the gravitational interaction and/or an anomalous
$U(1)$ symmetry and the magnitude of vacuum energy is not more than
of order $m_{3/2}^2 M^2$.}

\section{Introduction}
Superstring theories are powerful candidates for the unification
theory of all forces including gravity.
The supergravity theory (SUGRA) is effectively constructed from 
4-dimensional
(4D) string model using several methods.\cite{ST-SG,OrbSG,OrbSG2}
The structure of SUGRA is constrained
by gauge symmetries including an anomalous $U(1)$ symmetry \cite{ST-FI}
($U(1)_A$) and stringy symmetries 
such as duality.\cite{duality}
4D string models have several open questions.
\begin{enumerate}
\item There are thousands of effective theories 
corresponding to 4-D string models.
They have, in general, large gauge groups including $U(1)_A$
and many matter multiplets compared with those of the minimal 
supersymmetric standard model (MSSM).
We have not known how to select a realistic model from
stringy theoretical point of view yet.
The study on flat directions is important 
because effective theories have, in general, several flat directions 
in the SUSY limit.\cite{flat1}
Large gauge symmetries can break into smaller ones and extra 
matter fields can get massive through such flat directions.
Actually models with semi-realistic gauge groups and matter contents
have been constructed \cite{flat2} based on heterotic
$Z_3$ orbifold models.\footnote{Generic features of flat directions 
in $Z_{2n}$ orbifold models are studied.\cite{flat3}}

\item What is the origin of supersymmetry (SUSY) breaking?
Although intersting scenarios 
such as SUSY breaking mechanism due to gaugino condensation \cite{gaugino}
and Scherk-Schwarz mechanism \cite{SS}
have been proposed, realistic one has not been identified yet.

\item How is the vacuum expectation value (VEV) of dilaton
field $S$ stabilized?
It is difficult to realize the stabilization with a realistic VEV of $S$
using a K\"ahler potential at the tree level alone
without any conspiracy among several terms which 
appear in the superpotential.\cite{S-Stab}
A K\"ahler potential generally receives radiative
corrections as well as non-perturbative ones.
Such corrections may be sizable for the part related to
$S$.\cite{corr1,corr2}
\end{enumerate}

It is important to solve these enigmas
in order not only to understand the structure of more fundamental theory
at a high energy scale but also to know the complete SUSY particle spectrum
at the weak scale, but it is not an easy task
because of ignorance of the explicit forms of 
fully corrected total K\"ahler potential.
At present, it would be meaningful to get any information on SUSY
particle spectrum model-independently.\footnote{
The stability of $S$ and soft SUSY breaking parameters are discussed
in the dilaton dominated SUSY breaking scenario.\cite{Casas}}
On the second question, quite an interesting approach \cite{ST-soft}
has been adopted.
The formulae of soft SUSY breaking parameters have been derived in terms of
a few number of free parameters under the assumption that
the starting string model has the standard model gauge group 
and MSSM particle contents and
SUSY is broken by $F$-term condensations of the dilaton field
and/or moduli fields $M^i$.
Their phenomenological implications have been studied intensively.

The soft SUSY breaking parameters can be a powerful probe for high energy
physics beyond the MSSM because they are related to the
structure of the underlying theory.
For example, in SUSY grand unified theories,
the pattern of gauge symmetry breakdown can be specified
by checking certain sum rules among scalar masses.\cite{KMY1,KMY2,KT}
Specific relations among soft SUSY breaking
parameters can play an important role to probe 4D string 
models.\cite{KK,KKK}

In this paper, we study the magnitudes of soft SUSY breaking parameters
in heterotic string models with $U(1)_A$ and derive model-independent
predictions for them
without specifying both SUSY breaking mechanism
and the dilaton VEV fixing mechanism.
We assume that SUSY is broken by $F$-term condensation of $S$, 
$M^i$ and/or matter fields with non-vanishing $U(1)_A$ charge 
since the scenario based on $U(1)_A$ as a mediator
of SUSY breaking is also possible.\cite{DP,BD}
Soft SUSY breaking terms are calculated after 
SUSY breaking and flat direction breakings of gauge symmetries
in an effective SUGRA derived from 4D string model.
In particular, we make a comparison of magnitudes between
$D$-term contribution to scalar masses and $F$-term ones and 
a comparison of magnitudes among scalar masses, gaugino masses 
and $A$-parameters.
The features of our analysis are as follows.
The study is carried out in the framework of SUGRA 
model-independently,\footnote{The model-dependent analyses are 
carried out.\cite{DP,BD,MR,model-dep1,model-dep2}
Main results in this paper have been reported.\cite{model-indep}}
i.e., we do not specify SUSY breaking mechanism, extra matter contents, 
the structure of superpotential and
the form of K\"ahler potential related to $S$.
We treat all fields including $S$ and $M^i$ as dynamical fields. 

The paper is organized as follows. 
In section 2, we explain our starting effective SUGRA 
derived from heterotic string models with $U(1)_A$.
We study the magnitudes of soft SUSY breaking parameters 
model-independently in section 3.
Section 4 is devoted to conclusions and some comments.
In appendix, we explain a basic structure of SUGRA 
with assumptions of SUSY breaking.

\section{Heterotic string model with anomalous $U(1)$}

We explain our starting point and assumptions here.
The starting theory is an effective SUGRA 
on 4D heterotic string model.
The gauge group $G=G_{SM} \times U(1)_A$ originates from the 
breakdown of $E_8 \times E_8'$ gauge group
in 10D heterotic string theory.\footnote{
It is straightforward to apply our method to 4D string models with
a more generic gauge group 
$G=G'_{SM} \times U(1)^n \times U(1)_A \times H'$
where $G'_{SM}$ is a group which contains $G_{SM}$ as a subgroup,
$U(1)^n$ non-anomalous $U(1)$ symmetries, and $H'$ a direct product of
some non-abelian symmetries.}
Here $G_{SM}$ is a standard model gauge group
$SU(3)_C \times SU(2)_L \times U(1)_Y$ and $U(1)_A$ is an anomalous $U(1)$
symmetry.
The anomaly is canceled by the Green-Schwarz mechanism.\cite{GS}
Chiral multiplets are classified into two categories.
One is a set of $G_{SM}$ singlet fields which the dilaton field
$S$, the moduli fields $M^i$ and 
some of matter fields $\phi^m$ belong to.
The other one is a set of $G_{SM}$ non-singlet 
fields $\phi^k$.
We denote two types of matter multiplet as 
$\phi^\lambda = \{\phi^m, \phi^k\}$ .

The dilaton field $S$ transforms as 
$S \rightarrow S-i\delta_{GS}^{A} M \theta(x)$ under $U(1)_A$ 
with a space-time dependent parameter $\theta(x)$.
Here $M=M_{Pl}/\sqrt{8\pi}$ with $M_{Pl}$ being the Planck mass and
$\delta_{GS}^{A}$ is so-called Green-Schwarz coefficient
of $U(1)_A$ which is given by 
\begin{eqnarray}
  \delta_{GS}^{A} &=& {1 \over 96\pi^2}Tr Q^A 
        = {1 \over 96\pi^2} \sum_{\lambda} q^A_{\lambda} ,
\label{delta_GS}
\end{eqnarray}
where $Q^A$ is a $U(1)_A$ charge operator,
$q^A_{\lambda}$ is a $U(1)_A$ charge of $\phi^{\lambda}$ 
and the Kac-Moody level of $U(1)_A$ is rescaled as $k_A=1$.
We find $|\delta_{GS}^{A}/q^A_m| = O(10^{-1 \sim -2})$
in explicit models. \cite{KKK,KN}

The requirement of $U(1)_A$ gauge invariance yields the form 
of K\"ahler potential $K$ as,
\begin{eqnarray}
   K &=& K(S + {\bar S} + \delta_{GS}^{A} V_A, M^i, {\bar M}^i,
            {\bar \phi}_\mu e^{q^A_\mu V_A}, \phi^\lambda) 
\label{K-st}
\end{eqnarray}
up to the dependence on $G_{SM}$ vector multiplets.
We assume that derivatives of the K\"ahler potential $K$ with respect to
fields including 
moduli fields or matter fields are at most of order unity
in the units where $M$ is taken to be unity.
However we do not specify the magnitude of 
derivatives of $K$ by $S$ alone.
The VEVs of $S$ and $M^i$ are supposed to be
fixed non-vanishing values by some non-perturbative effects.
It is expected that the stabilization of $S$ is
due to the physics at the gravitational scale $M$ 
or at the lower scale than $M$.
Moreover we assume that the VEV of $K_S$ is much bigger than 
the weak scale, i.e., $O(m_{3/2}) \ll \langle K_S \rangle$.
The non-trivial transformation property of $S$
under $U(1)_A$ implies that $U(1)_A$ is broken down at some high energy
scale $M_I$.
The breaking scale of $U(1)_A$ is defined by 
$M_I \equiv |\langle \phi^m \rangle|$.

Hereafter we consider only the case with overall modulus 
field $T$ for simplicity.\footnote{
It is straightforward to apply our method to more complicated 
situations with multi-moduli fields.}
The K\"ahler potential is, in general, written by
\begin{eqnarray}
   K &=& K^{(S)}(S + {\bar S} + \delta_{GS}^{A} V_A) +
K^{(T)}(T + {\bar T}) + K^{(S, T)}
\nonumber\\ 
&~& + \sum_{\lambda, \mu} ( s_{\lambda}^{\mu}(S + {\bar S} 
+ \delta_{GS}^{A} V_A)
+ t_{\lambda}^{\mu}(T + {\bar T}) + u_{\lambda}^{{\mu}(S,T)} ) 
\phi^\lambda {\bar \phi}_\mu + \cdots 
\label{K-st2}
\end{eqnarray}
where $K^{(S, T)}$ and $u_{\lambda}^{{\mu}(S,T)}$ are mixing terms between
$S$ and $T$.
The magnitudes of $\langle K^{(S, T)} \rangle$, 
$\langle s_{\lambda}^{\mu} \rangle$ and 
$\langle u_{\lambda}^{{\mu}(S,T)} \rangle$
are assumed to be $O(\epsilon_1 M^2)$, $O(\epsilon_2)$ and $(\epsilon_3)$
where $\epsilon_n$'s ($n=1,2,3$) are model-dependent parameters 
whose orders are expected not to be more than one.\footnote{
The existence of $s_{\lambda}^{\mu} \phi^\lambda {\bar \phi}_\mu$
term in $K$
and its contribution to soft scalar masses are discussed in 
4D effective theory derived through the standard embedding 
from heterotic M-theory.\cite{het/M}}
We estimate the VEV of derivatives of $K$ 
in the form including $\epsilon_n$.
For example, $\langle K^\mu_{\lambda S} \rangle \leq O(\epsilon_p/M)$
($p=2,3$).
Our consideration is applicable to models in which some of $\phi^\lambda$
are composite fields made of original matter multiplets in string
models if the K\"ahler potential meets the above requirements.
Using the K\"ahler potential (\ref{K-st2}), $\hat{D}^A$ is given by
\begin{eqnarray}
  \hat{D}^A &=& -K_S \delta_{GS}^{A} M 
      + \sum_{\lambda, \mu} K_\lambda^\mu {\bar \phi}_\mu (q^A \phi)^\lambda
      + \cdots .
\label{D-st}
\end{eqnarray}
The breaking scale of $U(1)_A$ is estimated as
$M_I = O((\langle K_S \rangle \delta^A_{GS} M/q^A_m)^{1/2})$
from the requirement $\langle D^A \rangle \leq O(m_{3/2} M)$.
We require that $M_I$ should be equal to or be less than $M$,
and then we find that the VEV of $K_S$ has an upper bound such as
$\langle K_S \rangle \leq O(q^A_m M/\delta^A_{GS})$.

The $U(1)_A$ gauge boson mass squared $(M_{V}^{2})^A$ is given by
\begin{eqnarray}
 (M_{V}^{2})^A = 2 g_A^2 \left( \langle K^S_S \rangle (\delta_{GS}^A M)^2
            + \sum_{m, n} q^A_m q^A_n 
          \langle K^{n}_{m} \rangle 
          \langle \phi^m \rangle \langle {\bar \phi}_n \rangle \right)
\label{MV2A}
\end{eqnarray}
where $g_A$ is a $U(1)_A$ gauge coupling constant.
The magnitude of $(M_{V}^{2})^A/g_A^2$ is estimated as 
$Max(O(\langle K^S_S \rangle (\delta^A_{GS} M)^2), O(q^{A2}_m M_I^2))$.
We assume that the magnitude of $(M_{V}^{2})^A/g_A^2$ is 
$O(q^{A2}_m M_I^2)$.
It leads to the inequality 
$\langle K^S_S \rangle \leq O((q^{A}_m M_I/\delta^{A}_{GS} M)^2)$.

We discuss the relations among $\langle K_S \rangle$, 
$\langle K^S_S \rangle$ and $\langle K^S_{SS} \rangle$
using SUSY breaking conditions, i.e., SUSY is broken by the mixture of $S$,
$T$ and matter $F$-components such that
$\langle (K^S_S)^{1/2} F_S \rangle$, $\langle F_T \rangle$, 
$\langle F_m \rangle= O(m_{3/2} M)$,
and the stationary conditions of scalar potential.
The following relation is derived
\begin{eqnarray}
\langle K^S_S \rangle^{1/2} = O\left({\langle G_S \rangle \over M}\right)
= O\left({\langle K_S \rangle \over M} 
    + M {\langle W_S \rangle \over \langle W \rangle}\right)
\label{KSS-rel:app}
\end{eqnarray}
by the use of the definition (\ref{total-Kahler})
if the magnitude of $\langle K^S_S \rangle^{1/2}$ is much bigger than 
those of $\langle K^T_S \rangle$ and $\langle K^m_S \rangle$.
If $\langle K_S \rangle$ is bigger than 
$M^2 \langle W_S \rangle / \langle W \rangle$
and no cancellation happens among terms in 
$\langle K_S \rangle$ and $M^2 \langle W_S \rangle / \langle W \rangle$, 
we find that
$\langle K^S_S \rangle^{1/2} 
= O(\langle K_S \rangle / M) \leq O(q^A_m / \delta^{A}_{GS})$.
Further we can get the following relation among $\langle K_S \rangle$, 
$\langle K^S_S \rangle$ and $\langle K^S_{SS} \rangle$
from the stationary conditions (\ref{<VI>}) and (\ref{<VI2>}),
\begin{eqnarray}
{\langle K^S_{SS} \rangle \over \langle K^S_S \rangle} 
    = Max\left(O\left({\langle K^S_{S} \rangle 
      \over \langle K_S \rangle}\right),
          O\left({\langle K^S_{S} \rangle^{1/2} \over M}\right)\right).
\label{KSSS-rel:app}
\end{eqnarray}

\section{Magnitudes of soft SUSY breaking parameters}

We study the magnitudes of soft SUSY breaking parameters
in SUGRA from heterotic string model with $U(1)_A$.\footnote{
Based on the assumption that SUSY is broken by $F$-components of
$S$ and/or a moduli field, properties of
soft SUSY breaking scalar masses have been studied.\cite{N,KK}}
The formula of soft SUSY breaking scalar masses on $G_{SM}$ non-singlet
fields is given by \cite{KK}
\begin{eqnarray}
(m^2)^k_l &=& (m_{3/2}^2 + {\langle V_F \rangle \over M^2}) 
\langle K^k_l \rangle 
+ \langle {F}^{I} \rangle \langle {F}_{J} \rangle
(\langle R_{Il}^{Jk} \rangle + \langle X_{Il}^{Jk} \rangle) , \\
\langle R_{Il}^{Jk} \rangle &\equiv& 
\langle ({K}_{Il}^{I'} (K^{-1})^{J'}_{I'}
{K}^{kJ}_{J'} - {K}_{Il}^{Jk}) \rangle , \\
\langle X_{Il}^{Jk} \rangle &\equiv& 
 q^A_k (2 g_A^2/(M^2_V)^A) \langle ({\hat D}^A)_I^J \rangle
\langle K^k_l \rangle.
\end{eqnarray}
Here we neglect extra $F$-term contributions and so forth
since they are model-dependent.
The neglect of extra $F$-term contributions is justified 
if Yukawa couplings between heavy and light fields
are small enough and the $R$-parity violation is also tiny enough.
We have used Eq.(\ref{<VI2>}) to derive the part related to 
$D$-term contribution. 
Note that the last term in r.h.s. of Eq.(\ref{<VI2>}) is negligible
when $(M_{V}^{2})^A/g_A^2$ is much bigger than $m_{3/2}^2$.
Using the above mass formula, the magnitudes of 
$\langle R_{Il}^{Jk} \rangle$ and 
$\langle X_{Il}^{Jk} \rangle$ are estimated and given in Table 1.
Here we assume $q^A_k/q^A_m = O(1)$.

\begin{table}
\caption{The magnitudes of $\langle R_{Il}^{Jk} \rangle$ and 
$\langle X_{Il}^{Jk} \rangle$}
\begin{center}
\begin{tabular}{|c|l|l|}
\hline
$(I,J)$ & $\langle R_{Il}^{Jk} \rangle$ & $\langle X_{Il}^{Jk} \rangle$ \\
\hline\hline
$(S,S)$ & $O(\epsilon_p/M^2)$ & 
$Max(O(\langle K^S_{SS} \rangle/\langle K_{S} \rangle), 
O(\epsilon_p/M^2))$ \\ \hline
$(T,T)$ & $O(1/M^2)$ 
& $Max(O(\epsilon_1/(\langle K_{S} \rangle M)), 
O(1/M^2))$ \\ \hline
$(m,m)$ & $O(1/M^2)$ & $O(1/M_I^2)$ \\ \hline
$(S,T)$ & $O(\epsilon_p/M^2)$ 
& $Max(O(\epsilon_1/(\langle K_{S} \rangle M)), 
O(\epsilon_3/M^2))$ \\ \hline
$(S,m)$ & $O(\epsilon_p M_I/M^3)$
& $Max(O(\epsilon_p/(\langle K_{S} \rangle M)), 
O(\epsilon_p/(M M_I))$ \\ \hline
$(T,m)$ & $O(M_I/M^3)$
& $Max(O(\epsilon_3/(\langle K_{S} \rangle M)), 
O(1/(M M_I))$ \\ \hline
\end{tabular}
\end{center}
\end{table}

Now we obtain the following features on $(m^2)^k_l$.\\
(1) The order of magnitude of $\langle X_{Il}^{Jk} \rangle$ is
equal to or bigger than that of $\langle R_{Il}^{Jk} \rangle$
except for an off-diagonal part $(I,J)=(S,T)$.
Hence {\it the magnitude of $D$-term contribution is comparable to
or bigger than that of $F$-term contribution except for the universal part}
$(m_{3/2}^2 + \langle V_F \rangle/M^2)\langle K_l^k \rangle$.\\
(2) In case where the magnitude of $\langle F_m \rangle$ is bigger than
$O(m_{3/2} M_I)$ and $M > M_I$, 
we get the inequality $(m^2_D)_k \equiv \langle {F}^{I} \rangle 
\langle {F}_{J} \rangle \langle X_{Ik}^{Jk} \rangle  
> O(m_{3/2}^2)$ since the magnitude of 
$\langle \hat{D}^A \rangle$ is bigger than $O(m_{3/2}^2)$.\\
(3) In order to get the inequality $O((m^2_F)_k) > O((m^2_D)_k)$, the
following conditions must be satisfied simultaneously,
($(m^2_F)_k \equiv \langle {F}^{I} \rangle 
\langle {F}_{J} \rangle \langle R_{Ik}^{Jk} \rangle$)
\begin{eqnarray}
&~& \langle F_T \rangle, \langle F_m \rangle \ll O(m_{3/2} M) ,~~
\langle F_S \rangle = O\left({m_{3/2} M \over 
\langle K^S_S \rangle^{1/2}}\right) 
\nonumber \\
&~& {M^2 \langle K^S_{SS} \rangle \over \langle K_S \rangle
\langle K^S_S \rangle} < O(1) ,~~
{\epsilon_p \over \langle K^S_S \rangle} < O(1) ,~~(p=2,3)
\label{conditions}
\end{eqnarray}
unless an accidental cancellation among
terms in $\langle \hat{D}^A \rangle$ happens.
To fulfill the condition $\langle F_{T,m} \rangle \ll O(m_{3/2} M)$,
a cancellation among various terms including $\langle K_{I} \rangle$ and 
$\langle M^2 W_{I} / W \rangle$ is required. 
Note that the magnitudes of 
$\langle K_{T} \rangle$ and $\langle K_{m} \rangle$
are estimated as $O(M)$ and $O(M_I)$, respectively.

The gauge kinetic function is given by
\begin{eqnarray}
  f_{\alpha\beta} = k_\alpha {S \over M}\delta_{\alpha\beta}
  + \epsilon_\alpha  {T \over M}\delta_{\alpha\beta} 
+ f_{\alpha\beta}^{(m)}(\phi^\lambda) 
\label{f}
\end{eqnarray}
where $k_{\alpha}$'s are Kac-Moody levels
and $\epsilon_\alpha$ is a model-dependent parameter.\cite{epsilon}
The gauge coupling constants $g_\alpha$'s are related to the 
real part of gauge kinetic functions such that 
$g_\alpha^{2} = \langle Re f_{\alpha\alpha} \rangle^{-1}$.
The magnitudes of gaugino masses and $A$-parameters
in MSSM particles are estimated using the formulae
\begin{eqnarray}
M_a &=& \langle F^I \rangle \langle h_{aI} \rangle 
\label{Ma2} ,\\
\langle h_{aI} \rangle &\equiv& \langle Re f_a \rangle^{-1} 
\langle f_a,_I \rangle / 2\\
A_{kll'} &=& \langle F^I \rangle \langle a_{kll'I} \rangle 
\label{A2}, \\
\langle a_{kll'I} \rangle &\equiv& \langle f_{kll'},_I \rangle 
+ {\langle K_I \rangle \over M^2} \langle f_{kll'} \rangle
- \langle K_{(kI}^{I'} \rangle \langle (K^{-1})_{I'}^J \rangle 
\langle f_{Jll')} \rangle .
\end{eqnarray}
The result is given in Table 2.
Here we assume that $g_\alpha^{2} = O(1)$.

\begin{table}
\caption{The magnitudes of $\langle h_{aI} \rangle$ 
and $\langle a_{kll'I} \rangle$}
\begin{center}
\begin{tabular}{|c|l|l|}
\hline
$I$ & $\langle h_{aI} \rangle$ & $\langle a_{kll'I} \rangle$
\\ \hline\hline
$S$ & $O(1/M)$ & 
$Max(O(\langle K_{S} \rangle/M^2), O(\epsilon_p/M))$ \\ \hline
$T$ & $O(\epsilon_\alpha/M)$ & $O(1/M)$ \\ \hline
$m$ & $O(M_I/M^2)$ & $O(M_I/M^2)$ \\ \hline
\end{tabular}
\end{center}
\end{table}

In case that SUSY is broken by the mixture of $S$, $T$ and matter 
$F$-components such that
$\langle (K^S_S)^{1/2} F_S \rangle$, $\langle F_T \rangle$, 
$\langle F_m \rangle= O(m_{3/2} M)$ ,
we get the following relations among soft SUSY breaking parameters
\begin{eqnarray}
(m^2)_k &\geq& (m^2_D)_k = O\left(m_{3/2}^2 {M^2 \over M_I^2}\right) \geq
(A_{kll'})^2 = O(m_{3/2}^2) ,
\label{Mix-rel1}\\
(M_a)^2 &=& O(m_{3/2}^2) \cdot 
Max\left(O({\langle K^S_S \rangle}^{-1}),
O(\epsilon_\alpha^2), O\left({M_I^2 \over M^2}\right)\right) .
\label{Mix-rel2}
\end{eqnarray}

Finally we discuss three special cases of SUSY breaking scenario.

\begin{enumerate}
\item In the dilaton dominant SUSY breaking scenario
\begin{eqnarray}
\langle (K^S_S)^{1/2} F_S \rangle = O(m_{3/2} M) \gg 
\langle F_T \rangle, \langle F_m \rangle ,
\label{S-dom}
\end{eqnarray}
the magnitudes of soft SUSY breaking parameters are estimated as 
\begin{eqnarray}
&~&(m^2)_k = O(m_{3/2}^2) \cdot 
Max\left(O(1), O\left({M^2 \langle K^S_{SS} \rangle
\over \langle K^S_S \rangle \langle K_S \rangle}\right), 
O\left({\epsilon_p \over \langle K^S_S \rangle}\right)\right) , \nonumber \\
&~&M_a = O\left({m_{3/2} \over \langle K^S_S \rangle^{1/2}}\right) , ~~~
A_{kll'} = O(m_{3/2}) \cdot Max\left(O\left({\langle K_{S} \rangle 
\over M}\right), O(\epsilon_p)\right) . \nonumber
\end{eqnarray}
Hence we have a relation such that $O((m^2)_k)  \geq O((A_{kll'})^2)$.

Gauginos can be heavier than
scalar fields if $\langle K^S_S \rangle$ is small enough and
the inequality
$O(M^2 \langle K^S_{SS} \rangle) < O(\langle K_S \rangle)$
satisfied.
In this case, dangerous flavor changing neutral current 
(FCNC) effects from squark mass non-degeneracy are avoided 
because the radiative correction
due to gauginos dominates in scalar masses at the weak scale.
It is shown \cite{model-dep2} that gauginos are much lighter than
scalar fields from the requirement of 
the condition of vanishing vacuum energy 
in the SUGRA version of model \cite{BD} proposed 
by P.~Bin\'etruy and E.~Dudas.
We study it using the relations among the magnitudes of
$\langle K_S \rangle$, $\langle K^S_S \rangle$ and
$\langle K^S_{SS} \rangle$ discussed in section 2.
Let us consider a typical case with non-perturbative 
superpotential derived from SUSY breaking scenario by gaugino
condensations. 
The non-perturbative superpotential $W_{non}$ is, generally, given by
\begin{eqnarray}
W_{non} = \sum_i a_{i}(\phi^\lambda, T)
 \exp\left({-b_{i} S \over \delta^A_{GS} M}\right)
\label{W_{gaugino}}
\end{eqnarray}
where $a_{i}$'s are some functions of $\phi^\lambda$ and $T$, 
and $b_{i}$'s are model-dependent parameters of $O(q^A_m)$.
Using the second assumption, Eqs.(\ref{KSS-rel:app}) 
and (\ref{W_{gaugino}}), 
we get the relation
$\langle K^S_S \rangle^{1/2} 
= O(M \langle W_S \rangle / \langle W \rangle)
= O(q^A_m / \delta^{A}_{GS})$ 
if $O(\langle W_{non} \rangle) = O(\langle W \rangle)$.
This relation leads to $M_I = M$ from the third assumption
and the relation such that 
$M \langle K^S_S \rangle / \langle K_S \rangle 
= O(q^A_m / \delta^{A}_{GS})$.
We obtain the relation $M \langle K^S_{SS} \rangle / \langle K^S_S \rangle
= O(q^A_m  / \delta^{A}_{GS})$ using Eq.(\ref{KSSS-rel:app}).
Hence it is shown that the magnitude of gaugino masses is comparable to
or smaller than that of scalar masses.
It is necessary to relax some of assumptions in order to
have the SUSY spectrum at $M_I$ such that gauginos are much heavier than 
scalar fields.
A possibility is to introduce a constant term in $W$ such that
$\langle W_{non} \rangle/\langle W \rangle 
= O((M_I / M)^2)$ and to cancel a leading term in 
$\langle K_S \rangle$ by $M^2 \langle W_S \rangle / \langle W \rangle$.

\item In the moduli dominant SUSY breaking scenario
\begin{eqnarray}
\langle F_T \rangle = O(m_{3/2} M) \gg 
\langle (K^S_S)^{1/2} F_S \rangle, \langle F_m \rangle ,
\label{T-dom}
\end{eqnarray}
the magnitudes of soft SUSY breaking parameters are estimated as 
\begin{eqnarray}
&~&(m^2)_k = O(m_{3/2}^2) \cdot Max\left(O(1), O\left({\epsilon_1 M \over
\langle K_S \rangle}\right)\right) ,\nonumber \\
&~&M_a = O(\epsilon_\alpha m_{3/2}) , ~~~
A_{klm} = O(m_{3/2}) . \nonumber
\end{eqnarray}
Hence we have a relation such that
$O((m^2)_k) \geq O((A_{kll'})^2) \geq O((M_a)^2)$.
The magnitude of $\mu_{TT}$ is estimated as $\mu_{TT} = O(m_{3/2})$.

\item In the matter dominant SUSY breaking scenario
\begin{eqnarray}
\langle F_m \rangle = O(m_{3/2} M) \gg 
\langle (K^S_S)^{1/2} F_S \rangle, \langle F_T \rangle ,
\label{matter-dom}
\end{eqnarray}
the magnitudes of soft SUSY breaking parameters are estimated as 
\begin{eqnarray}
(m^2)_k &=& O\left(m_{3/2}^2 {M^2 \over M_I^2}\right) , ~~~
M_a, A_{kll'} = O\left(m_{3/2} {M_I \over M}\right) . \nonumber 
\end{eqnarray}
The relation $(m^2)_k \gg O((M_a)^2) = O((A_{kll'})^2)$
is derived when $M \gg M_I$.
The magnitude of $\mu_{mn}$ is estimated as 
$\mu_{mn} = O(m_{3/2} M/M_I)$.
\end{enumerate}

\section{Conclusions}

We have studied the magnitudes of soft SUSY breaking parameters
in heterotic string models with $G_{SM} \times U(1)_A$,
which originates from the breakdown of $E_8 \times E'_8$, and derive 
model-independent predictions for them
without specifying both SUSY breaking mechanism
and the dilaton VEV fixing mechanism.
In particular, we have made a comparison of magnitudes between
$D$-term contribution to scalar masses and $F$-term ones and 
a comparison of magnitudes among scalar masses, gaugino masses 
and $A$-parameters
under the condition that $O(m_{3/2}) \ll \langle K_S \rangle \leq
O(q^A_m M/\delta^A_{GS})$, $(M_V^2)^A/g_A^2 = O(q^{A2}_m M_I^2)$
and $\langle V \rangle \leq O(m_{3/2}^2 M^2)$.
The order of magnitude of $D$-term contribution of $U(1)_A$
to scalar masses is comparable to
or bigger than that of $F$-term contribution $\langle F^I \rangle
\langle F_J \rangle \langle R_{Il}^{Jk} \rangle$
except for the universal part 
$(m_{3/2}^2 + \langle V_F \rangle/M^2)\langle K_l^k \rangle$.
If the magnitude of $F$-term condensation of matter fields 
$\langle F_m \rangle$ is bigger than
$O(m_{3/2} M_I)$, the magnitude of $D$-term contribution $(m_D^2)_k$ 
is bigger than $O(m_{3/2}^2)$. 
In general, it is difficult to realize the inequality 
$O((m^2_D)_k) < O((m^2_F)_k)$ unless conditions such as 
Eq.(\ref{conditions}) are fulfilled.
We have also discussed relations among soft SUSY breaking
parameters in three special scenarios on SUSY breaking,
i.e., dilaton dominant SUSY breaking scenario,
moduli dominant SUSY breaking scenario 
and matter dominant SUSY breaking scenario.
In the same way, we can estimate the magnitude of $D$-term contribution
to scalar masses for non-anomalous diagonal charges.
The magnitudes of $\langle X_{Il}^{Jk} \rangle$ is given by
$O(\epsilon_p/M^2)$, $O(1/M^2)$, $O(1/M_B^2)$, $O(\epsilon_3/M^2)$,
$O(\epsilon_p/(M M_B))$ and $O(1/(M M_B))$
for $(I,J)$ $=$ $(S,S)$, $(T,T)$, $(m,m)$, $(S,T)$, $(S,m)$ and $(T,m)$
where $M_B$ is a breaking scale of extra gauge symmetry.

The $D$-term contribution to scalar masses with different broken
charges as well as the $F$-term contribution from the difference among
modular weights can destroy universality among scalar masses.
The non-degeneracy among squark masses of first and second families
endangers the discussion of the suppression
of FCNC process.
On the other hand, the difference among broken charges
is crucial for the scenario of fermion mass hierarchy 
generation.\cite{texture}
It seems to be difficult to make two discussions compatible.
There are several way outs.
The first one is to construct a model that the fermion mass hierarchy is 
generated due to non-anomalous $U(1)$ symmetries.
In the model, $D$-term contributions of non-anomalous $U(1)$ symmetries
vanish in the dilaton dominant SUSY breaking case
and it is supposed that anomalies from contributions of
the MSSM matter fields are canceled out 
by an addition of extra matter fields.
The second one is to use ``stringy'' symmetries 
for fermion mass generation in the situation with degenerate 
soft scalar masses.\cite{texture2}

Finally we give a comment on moduli problem.\cite{cosmo}
If the masses of dilaton or moduli fields are of order of the weak scale,
the standard nucleosynthesis should be modified
because of a huge amount of entropy production.
The dilaton field does not cause dangerous contributions
in the case with $\langle (K^S_S)^{1/2} F_S \rangle = O(m_{3/2} M)$
if the magnitude of $\langle K_S^S \rangle$
is small enough,\footnote{
This possibility has been pointed out
in the last reference in 13.}
because the magnitudes of $(m_F^2)_S$ is
given by $O(m_{3/2}^2/\langle K_S^S \rangle^2)$.

\section*{Acknowledgements}
The author is grateful to T.~Kobayashi, H.~Nakano, H.P.~Nilles and 
M.~Yamaguchi for useful discussions.

\section*{Appendix}

We review the scalar potential part in SUGRA.\cite{SUGRA,JKY,Kawa1}  
It is specified by two functions, the total K\"ahler
potential $G(\phi, \bar \phi)$ and the gauge kinetic function
$f_{\alpha \beta}(\phi)$ with $\alpha$, $\beta$  being indices of 
the adjoint representation of the gauge group.   
The former is a sum of the K\"ahler potential $K(\phi, \bar \phi)$ 
and (the logarithm of) the superpotential $W(\phi)$
\begin{equation}
   G(\phi, \bar \phi)=K(\phi, \bar \phi) +M^{2}\ln |W (\phi) /M^{3}|^2 .
\label{total-Kahler}
\end{equation}
We have denoted scalar fields in the chiral multiplets by $\phi^I$
and their complex conjugate by $\bar \phi_J$.  
The scalar potential is given by
\begin{eqnarray}
   V &=& M^{2}e^{G/M^{2}} (G_I (G^{-1})^I_J G^J-3M^{2})
 + \frac{1}{2} (Re f^{-1})_{\alpha \beta} \hat D^{\alpha} \hat D^{\beta}
\label{scalar-potential}
\end{eqnarray}
where
\begin{eqnarray}
\hat D^\alpha =  G_I ( T^\alpha \phi)^I = (\bar \phi T^\alpha)_J G^J.
\label{hatD}
\end{eqnarray}
Here $G_I=\partial G/\partial \phi^I$, 
$G^J=\partial G/\partial \bar \phi_J$ etc, and
$T^\alpha$ are gauge transformation generators. 
Also in the above, $(Re f^{-1})_{\alpha \beta}$ and  
$(G^{-1})^I_J$ are the inverse matrices of $Re f_{\alpha \beta}$ and  
$G^I_J$, respectively, and a summation over $\alpha$,... and  $I$,... is
understood.  
The last equality in Eq.(\ref{hatD}) comes from the
gauge invariance of the total K\"ahler potential.
The $F$-auxiliary fields of the chiral multiplets and 
the $D$-auxiliary fields of the vector multiplets are given by
\begin{equation}
   F^I =Me^{G/2M^{2}} (G^{-1})^I_J G^J ,  ~~~~ 
   D^{\alpha} = (Re f^{-1})_{\alpha\beta} \hat{D}^{\beta} ,
\end{equation}
respectively.
Using $F^I$ and $D^\alpha$, the scalar potential is rewritten down by
\begin{eqnarray}
   V &=& V_F + V_D , \nonumber \\
  V_F &\equiv& F_I K^I_J F^J - 3M^{4} e^{G/M^{2}}  ,
\label{VF}\\
   V_D &\equiv& \frac{1}{2} Re f_{\alpha \beta} D^{\alpha} D^{\beta}.
\label{scalar-potential 2}
\end{eqnarray}

Let us next summarize our assumptions on SUSY breaking.  
The gravitino mass $m_{3/2}$ is given by
\begin{equation}
   m_{3/2}= \langle Me^{G/2M^{2}} \rangle
\label{gravitino}
\end{equation}
where $\langle \cdots \rangle$ denotes the VEV.  
As a phase convention, it is taken to be real.  
We identify the gravitino mass with the weak scale in most cases.
It is assumed that SUSY is spontaneously broken by some $F$-term condensations
($\langle F \rangle \neq 0$) for singlet fields under 
the standard model gauge group
and/or some $D$-term condensations ($\langle D \rangle \neq 0$)
for broken gauge symmetries.
We require that the VEVs of $F^I$ and $D^{\alpha}$ should satisfy
\begin{eqnarray}
   \langle (F_I K^I_J F^J)^{1/2} \rangle \leq O(m_{3/2}M) ,
~~~ \langle D^\alpha \rangle \leq O(m_{3/2}M) 
\label{VEV}
\end{eqnarray}
for each pair $(I,J)$.
Note that we allow the non-zero vacuum energy $\langle V \rangle$
of order $m_{3/2}^2 M^2$ at this level, which could be canceled
by quantum corrections.

In order to discuss the magnitudes of several quantities, it is
necessary to see consequences of the stationary condition 
$\langle \partial V /\partial \phi^I \rangle =0$.   From
Eq.(\ref{scalar-potential}),  we find
\begin{eqnarray}
 \partial V /\partial \phi^I &=& G_I \left( {V_F \over M^2}
   + M^{2}e^{G/M^{2}} \right) + M e^{G/2M^{2}} G_{IJ} F^J \nonumber \\
 &~& - F_{I'} G^{I'}_{J'I} F^{J'} 
     - \frac{1}{2} (Re f_{\alpha \beta}),_I D^\alpha D^\beta
\nonumber \\
 &~& + D^\alpha (\bar \phi T^\alpha )_J G^J_I .
\label{VI}
\end{eqnarray}
Taking its VEV and using the stationary condition, we derive the formula
\begin{eqnarray}
 m_{3/2} \langle G_{IJ} \rangle \langle F^J \rangle &=& 
 - \langle G_I \rangle 
  \left( {\langle V_F \rangle \over M^2} + m_{3/2}^2 \right) 
 + \langle F_{I'} \rangle \langle G^{I'}_{J'I} \rangle 
   \langle F^{J'} \rangle
\nonumber \\
&~& + \frac{1}{2} \langle (Re f_{\alpha \beta}),_I \rangle 
   \langle D^\alpha \rangle \langle D^\beta \rangle 
    - \langle D^\alpha \rangle \langle (\bar \phi T^\alpha )_J \rangle
   \langle G^J_I \rangle .
\label{<VI>}
\end{eqnarray}
We can estimate the magnitude of SUSY mass parameter 
$\mu_{IJ} \equiv m_{3/2} (\langle G_{IJ} \rangle + \langle G_{I} \rangle
\langle G_{J} \rangle/M^2 - \langle G_{I'} (G^{-1})^{I'}_{J'} 
G^{J'}_{IJ} \rangle)$ using Eq.(\ref{<VI>}).
By multiplying $(T^\alpha \phi)^I$ to Eq.(\ref{VI}), 
a heavy-real direction is projected on.  
Using the identities derived from the gauge
invariance of the total K\"ahler potential
\begin{eqnarray}
   & & G_{IJ}(T^\alpha \phi)^J + G_J (T^\alpha )_I^J
   - K^J_I(\bar \phi T^\alpha)_J = 0,
\\
   & & K_{IJ'}^J (T^\alpha \phi)^{J'} + K^J_{J'} (T^\alpha)^{J'}_I
       -[G^{J'} (\bar \phi T^\alpha )_{J'}]_I^J = 0,
\end{eqnarray}
we obtain
\begin{eqnarray}
  {\partial V \over \partial \phi^I} (T^\alpha \phi)^I &=&
      \left( {V_F \over M^2} + 2 M^2 e^{G/M^2} \right) \hat D^\alpha
       - F_I F^J (\hat{D}^{\alpha})^I_J 
\nonumber \\
 &~& - \frac{1}{2} (Re f_{\beta \gamma}),_I (T^\alpha \phi)^I
       D^\beta D^\gamma
      + (\bar \phi T^\beta)_J G^J_I (T^\alpha \phi)^I D^\beta .
\label{VI2}
\end{eqnarray}
Taking its VEV and using the stationary condition, 
we derive the formula
\begin{eqnarray}
&~& \left( {(M_V^2)^{\alpha\beta} \over 2g_\alpha g_\beta} 
+ \left({\langle V_F \rangle \over M^2} + 2 m_{3/2}^2\right)
 \langle Re f_{\alpha\beta} \rangle \right) \langle D^\beta \rangle
       = \langle F_I \rangle \langle F^J \rangle 
         \langle (\hat{D}^{\alpha})^I_J \rangle \nonumber \\
&~& ~~~~~~~~~~~~~~~~~~~~~~~~~~~~ 
+ \frac{1}{2} \langle (Re f_{\beta \gamma}),_I \rangle 
   \langle (T^\alpha \phi)^I \rangle 
   \langle D^\beta \rangle \langle D^\gamma \rangle 
\label{<VI2>}
\end{eqnarray}
where $(M_{V}^{2})^{\alpha \beta}=
2g_\alpha g_\beta \langle (\bar \phi T^\beta)_J K^J_I 
(T^\alpha \phi)^I \rangle$
is the mass matrix of the gauge bosons and
$g_\alpha$ and $g_\beta$ are the gauge coupling constants.  
Using Eq.(\ref{<VI2>}), we can estimate the magnitude of
$D$-term condensations $\langle D^\beta \rangle$.
The formula of $\langle D^\beta \rangle$ is given by
\begin{eqnarray}
&~& \langle D^\beta \rangle = \langle F_I \rangle \langle F^J \rangle 
        {2g_\alpha g_\beta \over (M_V^2)^{\alpha\beta}}
 \langle (\hat{D}^{\alpha})^I_J \rangle
\label{<Dbeta>}
\end{eqnarray}
in case where 
$(M_V^2)^{\alpha\beta}/g_\alpha g_\beta \gg O(m_{3/2}^2)$.

\end{document}